\documentclass{intech07}

\booktitle{Will-be-set-by-IN-TECH}%

\chaptertitle{The Concept of Temperature in \\the Modern Physics} % You know your chapter title?

\authors{Dmitrii Tayurskii$^1$ and Alain Le M\'{e}haut\'{e}$^2$}
\affiliation{$^1$Kazan Federal University\\
$^2$Institute International Franco-Qu\'{e}b\'{e}cois}
\country{$^1$Russia\\
$^2$France}

\begin{document}

\maketitle

\section{Introduction }
Thermodynamics, or thermodynamic method, has been invented as a way to describe the exchange processes of energy and matter occurring at the molecular and atomic levels without considering the details of molecular motion. The very beginning of it has been inspired by steam engines efficiency and the initial context of thermodynamics concerned the macroscopic systems in equilibrium and/or quasi-equilibrium states. Statistical physics has been appeared as an attempt to describes the same processes with taking into account atomic hypothesis and atomic dynamic parameters, i.e. the velocities and the positions of atoms. The probabilistic language appeared to be very suitable for these purposes. But again the studying has been restricted by the macroscopic systems in quasi-equilibrium states mostly because of using variational principles like the maximum entropy principle.

Today the relevance of thermodynamic formalism and applicability of statistical physics are questionable when the nanosystems, the systems far from equilibrium and the systems with strong interactions begin to be studied. First of all the problems appear for the definition of the temperature that is the key concept in the formalism of thermodynamics and statistical physics. The assumptions used to define the temperature have to be treated very carefully in the cases of nanosystems, systems with strong interactions and other complex systems\footnote{During the preparation of the present paper the very detailed book has been published \citep{biro2011} where the basic concepts at the very heart of statistical physics are presented and their challenges in high energy physics are discussed}.

The goal of the present paper is to remind the main conditions which has to be satisfied to introduce the physical quantity ``temperature'' and to discuss the possibilities for introducing the temperature in complex systems. As an example the model system - the system of interacting spins at external magnetic field - will be used to demonstrate the advantages and restrictions of using ``spin temperature'' concept.

\section{What is the temperature?}
The thermodynamic or statistical definition of the temperature can be found in any standard physics textbook. Here we reproduce briefly these procedures with emphasizing some points which are usually considered as the given.
As far as the exchange processes of energy and matter are the subject of investigations two macroscopic systems in equilibrium are usually considered. If one allows only energy exchange between these two systems the equilibrium means the equality of some physical parameter in this case. We can call this parameter as a temperature. But how we can measure and/or calculate this parameter? It is necessary to mention here that the key-point in the definition of the temperature is the existence of the so-called thermal equilibrium between two systems.

In the framework the phenomenological approach - thermodynamics - the temperature is measured by the monitoring other physical parameters (expansion coefficient, resistivity, voltage, capacity etc.). That is why there are so many different kinds of thermometers. The procedure of temperature measurements consists of the thermal contact (energy exchange) between the system under consideration and a thermometric body the physical state of which is monitored. This thermometric body should be as small as possible in order to do not disturb the state of the system during measurement. In the realm of very small systems such a procedure is rather questionable. What the size should be for the thermometer to measure, for example, the temperature of a nanosystem? Should the thermometric body be an atom or elementary particle in this case? But the states of atoms and elementary particles are essential quantum ones and can not be changed continuously.
The excellent treatment of the more sophisticated measurements of temperature (spectral temperature and radiation temperature) the reader can find in the very recent book \citep{biro2011}.

The simplest way to determine the temperature in statistical physics is based on the consideration of possible microscopic states for the given macrostate which is determined by the energy of system or other constraints. The number of possible microscopic states $\Omega(E)$ is extremely fast rising function of the energy of system $E$. For two systems being in the thermal contact the total number of possible microscopic states is given usually by the following product:

\begin{equation}\label{product}
    \Omega_{12}=\Omega(E_1)\cdot\Omega(E_2).
\end{equation}

An implicit assumption in the notion of thermal contact is that the system-system interaction is vanishingly small, so that the total energy $E$ is simply given by

\begin{equation}\label{energy}
    E=E_{1}+E_{2}=const.
\end{equation}

\begin{figure}[htb]	
\centering
\includegraphics[width=9cm]{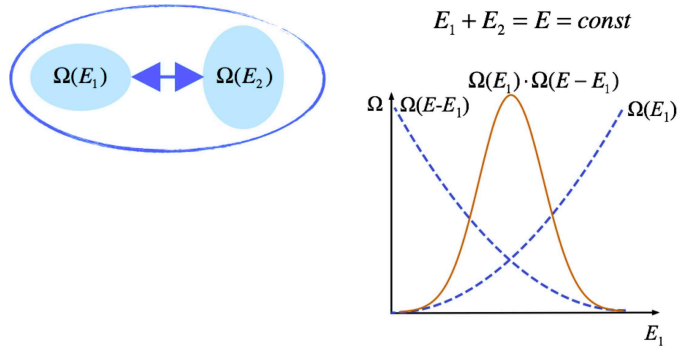}
\caption{Two systems in the thermal contact and the number of possible microscopic states for them and for the joint system in dependence on energy. Note that in the definition of thermal contact the energy of interaction between systems is vanishingly small, so the total energy is just $E=E_1+E_2$.}\label{maximum}
\end{figure}

The product $\Omega(E_1)\cdot\Omega(E-E_1)$ shows the very sharp maximum (see Figure \ref{maximum}) and it is more convenient to study the extremal conditions for the logarithm  $ln(\Omega(E_1)\cdot\Omega(E-E_1))$ from which on immediately get the definition of the statistical temperature:

\begin{equation}\label{stattemp}
    \frac{\partial ln(\Omega(E_1))}{\partial E_1}= \frac{\partial ln(\Omega(E_2))}{\partial E_2}=\beta=\frac{1}{k_BT},
\end{equation}

where $k_B$ is the Boltzmann's constant and $T$ denotes the absolute thermodynamics temperature.
In fact the taking of logarithm leads us to some additive quantity, and it is the property which is carried by the Boltzmann definition of entropy:

\begin{equation}\label{Bentropy}
    S=k_{B}ln\Omega(E).
\end{equation}

Considering the system in the contact with the thermal bath (thermal reservoir) the same assumption about neglecting system-bath interaction leads to the existence of canonical (Gibbs) distribution for the probabilities to find the system in the state with energy $E_{\alpha}$:

\begin{equation}\label{Gibbs}
    P(E_{\alpha})\propto exp(-\beta E_{\alpha}).
\end{equation}

We have seen that all obtained results are valid only if one neglects by the system-system or system-bath interactions. In fact such neglecting proved to be the consequence of the so-called thermodynamic limit which is reached as the number of particles (atoms or molecules) in a system, $N$, approaches infinity. For the most systems in the thermodynamic limit the macroscopic extensive variables (energy, entropy, volume) possess the property of additivity like Equation \ref{energy}. It is necessary to point out here that the possibility to describe the thermodynamic behavior of the system under consideration  by the statistical physics methods as $N$ tends to infinity is not automatically granted but depends crucially of the nature of the system. It has been shown many times that the Gibbs canonical ensemble is valid only for sufficiently short range interactions and there are examples - self-gravitating systems, unscreened Coulomb systems -  for which the assumed additivity postulate is violated.

Also it should be mentioned that it is possible to give an alternative definition of the thermodynamic limit \citep{biro2011}. If a system is so large that it itself can serve as a perfect thermal reservoir (bath) for its smaller parts, then one can consider this system as being in the thermodynamic limit. This definition is not restricted to large volume and large particle number. But again here the energy of spin-bath interactions is neglected and, in practice, such definition also implies the existence of additive quantities.

The more rigorous definition of temperature in statistical physics is based on the maximization of entropy for the system composed from two subsystems (like we consider above) while the energy, volume, particle number etc. are composed from the corresponding subsystems value (see, for example, \citep{biro2011}). Mathematically in denotes that one should look for the maximum of the entropy of the system

\begin{equation}\label{maxentr}
   S(E,V,N,\ldots)=max
\end{equation}

when the following equations are satisfied

\begin{equation}\label{cond}
\begin{array}{rcl}
 E=E_1\oplus E_2 \\
 V=V_1\oplus V_2 \\
 N=N_1\oplus N_2.
\end{array}
\end{equation}

Strictly speaking the composition law $\oplus$ in Equations \ref{cond} is not restricted only by addition. But finding the maximum of entropy \ref{maxentr}

\begin{equation}\label{maxentr}
   dS(E,V,N,\ldots)=\left(\frac{\partial S}{\partial E_1}dE_1+\frac{\partial S}{\partial E_2}dE_2\right)+\left(\frac{\partial S}{\partial V_1}dV_1+\frac{\partial S}{\partial V_2}dV_2\right)+\left(\frac{\partial S}{\partial N_1}dN_1+\frac{\partial S}{\partial N_2}dN_2\right)+\ldots=0
\end{equation}

 when the total energy, volume and particle number are fixed, i.e.

\begin{equation}\label{energyfix}
   dE=\left(\frac{\partial E}{\partial E_1}dE_1+\frac{\partial E}{\partial E_2}dE_2\right)=0
\end{equation}

  \begin{equation}\label{volumefix}
   dV=\left(\frac{\partial V}{\partial V_1}dV_1+\frac{\partial V}{\partial V_2}dV_2\right)=0
\end{equation}

\begin{equation}\label{numberfix}
   dN=\left(\frac{\partial N}{\partial N_1}dN_1+\frac{\partial N}{\partial N_2}dN_2\right)=0
\end{equation}

(please note, that here energy, volume and particle number are considered as independent variables!!!) one gets the following expressions:

 \begin{equation}\label{maxentr1}
   \left(\frac{\partial S}{\partial E_1}\frac{1}{\frac{\partial E}{\partial E_1}}\right)=\left(\frac{\partial S}{\partial E_2}\frac{1}{\frac{\partial E}{\partial E_2}}\right)
\end{equation}

 \begin{equation}\label{maxentr2}
   \left(\frac{\partial S}{\partial V_1}\frac{1}{\frac{\partial V}{\partial V_1}}\right)=\left(\frac{\partial S}{\partial V_2}\frac{1}{\frac{\partial V}{\partial V_2}}\right)
\end{equation}

 \begin{equation}\label{maxentr3}
   \left(\frac{\partial S}{\partial N_1}\frac{1}{\frac{\partial N}{\partial N_1}}\right)=\left(\frac{\partial S}{\partial N_2}\frac{1}{\frac{\partial N}{\partial N_2}}\right)
\end{equation}

The left (right) part of Equation \ref{maxentr1} will depends on the quantities describing subsystem 1 (2) only in the case, when the simplest addition is taken in \ref{cond} as the composition law. So again we conclude that the rigorous definition of the temperature can be done only for the extensive systems in thermodynamic limit.

Now we will demonstrate that the thermodynamic limit does not exist for the systems with strong enough interactions and how this limit can be restored by some averaging procedure. We consider a $N$-particle system describing by the following Hamiltonian:

\begin{equation}\label{Ham}
    \mathcal{H}=\sum_{i=1}^N \mathcal H_{0i}+\sum_{i<j}^N \mathcal U_{ij},
\end{equation}

where $\mathcal H_{0i}$ is the Hamiltonian of free particle and $\mathcal U_{ij}$ describes the interaction between $i$-th and $j$-th particles. It is obviously that because of second term in Equation \ref{Ham} the energy of system is not additive and the system can not be trivially divided into two- or more independent subsystems. By increasing the number of particles in the system in $n$ times the energy of systems is not increased in $n$ times too. The conditions of thermodynamic limit are violated here. But if the interaction between particles $\mathcal U_{ij}$ is a short-range one and a rather small the concept of the mean field can be introduced, so the Hamiltonian \ref{Ham} can be re-written as:

\begin{equation}\label{HamMF}
    \mathcal{H}=\sum_{i=1}^N \mathcal H_{0i}+\sum_{i=1}^N\langle \mathcal{U}_{ij} \rangle _{j}.
\end{equation}

So the energy becomes additive and the conditions of thermodynamic limit are satisfied.

We note here that for nanoscale systems in which the contribution to their energy (or other quantity) from the surface atoms is comparable with that from the bulk volume atoms the non-additivity exists forever. The surface energy is increased in $n^{2/3}$ times when one increases the number of particles in $n$ times. That is why the question about the definition of temperature for nanoscale system is very intriguing not only from the point of view of its physical measurements but also from the estimations for the minimal length scale on which this intensive quantity exists \citep{hartmann04}.

Last decades the non-extensive thermodynamics has being developed  to describe the properties of systems where the thermodynamic limit conditions are violated. It is not a purpose of this paper to give one more review of non-extensive thermodynamics, its methods and formalism. The reader can found it in numerous papers, reviews and books (see, for example, \citep{tsallis09,abe07,gellmann04,abe2001}). Here we would like to underline only that the definition of temperature is very close related to the existence of thermal equilibrium and is very sensitive to the thermodynamic limit conditions, so a researcher should be very careful in the prescribing the meaning of temperature to a Lagrange multiplier when entropy maximum is looked for.

In the next part of this work we will show how the thermodynamic formalism can be applied for the system of spins in an external magnetic field and will discuss the existence of two spin temperatures for one spin system. It is a real good example when one has to remember all conditions being used to

\section{Spin temperature}
An important progress in the description of the behavior of spin degrees of freedom in solids was reached with the use of thermodynamic approach. The relation between electron and nuclear magnetism on the one hand and thermodynamics on the other hand was first established by Casimir and Du Pre \citep{casimir38}. They introduced the concept of spin temperature for a system of non-interacting spins in an external magnetic field. But for a long time spin temperature was being considered an elegant theoretical representation only. Further investigations of spin thermodynamics showed that spin temperature can be experimentally measured and its change is connected with the transfer of heat and change of entropy. For example in the framework of spin temperature concept Bloembergen et al. \citep{bloemb48,bloemb59} built the classical theory of saturation and cross-relaxation in spin systems. The excellent account of spin temperature concept the reader can find in \citep{abragam94,goldman70,atsarkin72}.

In the classical theory of Bloembergen et al.  \citep{bloemb48} the Zeeman levels are considered as being infinitely sharp thus neglecting the broadening due to spin-spin interactions. This is only justified in cases of liquids and gases, where the rapid motion of the atoms or molecules averages the spin-spin interactions to zero and the spins can be considered as being independent of each other. However in solids the spin-spin interactions are normally so strong that the whole ensemble of spins acts as a collective system with many degrees of freedom.
The next main step in the understanding of spin thermodynamics in solids was made by Shaposhnikov \citep{shap47,shap48,shap49}, whose works were far in advance of the experimental possibilities of their verifying. In these little known works Shaposhnikov pointed out the significance of taking into account the interactions inside spin systems. As long as the time taken for internal thermal equilibrium to be established is finite then to describe any state of spin system one needs to determine two thermodynamic coordinates, which are magnetization and spin temperature. As distinct from the theory of Casimir and Du Pr\'e these coordinates are not connected between each other by Curie law. Moreover, the spin temperature and total energy of spin system were proven to be just thermodynamically conjugated variables, and the magnetization characterizes a state of spin system in an external magnetic field.

Independently in 1955 to investigate the saturation in system of interacting spins Redfield \citep{redf1955} introduced   the hypothesis that under strong saturation the whole spin system stayed in internal equilibrium, thus permitting its description by one single temperature. The problem for an arbitrary degree of saturation was solved in 1961 by Provotorov who developed Shaposhnikov's ideas about taking into account spin- spin interactions. In well-known works \citep{prov62a,prov62b,prov63} Provotorov showed that under some conditions an energy of spin-spin interactions that are small compared with the interaction with strong external magnetic field can be extracted into a separate thermodynamic subsystem called by the reservoir of spin-spin interactions. This thermodynamic subsystem has its own temperature different from the temperature of spin system in an external magnetic field determined in Gasimir and Du Pr\'e theory. Therefore according to the Provotorov's theory any state of spin system can be described by two temperatures. The concept of two temperatures turned out to be fruitful and it was confirmed experimentally in electron as well as in nuclear magnetism. This concept led to the revision of some representations in the theory of magnetic resonance and relaxation and to the prediction of a number of unexpected physical effects.

Today the two-temperature formalism being unusual from the point of view of statistical thermodynamics forms the basic framework for the theory of magnetic resonance in solids. But because of conceptual and mathematical difficulties the Provotorov's theory was well developed only in the so-called high-temperature approximation when the heat energy of spin greatly exceeds the energy of spin in external magnetic field.

The attempts to extend the theory towards low temperatures (the energy of spin in external field is more than heat energy) demonstrated the principal difficulties in the choosing of thermodynamic variables and in the understanding of energy redistribution inside spin-system (see, for review  \citep{tayur89,tayur90} and references therein). But there are many experiments at low temperatures to interpret of which it is necessary to have a theory describing the spin thermodynamics and kinetics in this case. Among them we point out dynamic polarization experiments \citep{abragam82}, nuclear ordering in solids \cite{loun1989,goldman1991}.
The object of this paper is to give a description of some theoretical approaches to the studying of spin-system in solids at low temperatures.
In the next section we will summarize the concept of spin temperature, as far as it is very important for the understanding of low temperature thermodynamics of spin-system in solids.

\subsection{High and low temperatures}
If an electron (or nuclear) spin $S$ is subjected to a magnetic field $H_0$ in the direction $z$    then the Zeeman interaction

\begin{equation}\label{Zeeman}
\mathcal{H}_{z}=\omega_0S^z
\end{equation}

establishes a set of $2S+1$ sublevels with energy $E_m= m\omega_0$
(here and further the units $\hbar=1$ are used ); $\omega_0=\gamma H_0$ is the?Larmor frequency, $\gamma$ is the gyromagnetic ratio, $m$ is the magnetic quantum number ( $m =-S,-S+1,\ldots S-1,S$). In Equation \ref{Zeeman} one must take the negative sign if $S$ is a nuclear spin. For an
ensemble of $N$ identical spins $S=1/2$ we can introduce the Zeeman level populations $n_{+}$ and $n_{-}$ ( $n_{+}+n_{-}=N$ ), where $n_{+}$ is the number of spins in the state $m=+1/2$ and $n_{-}$ is the number of spins with $m=-1/2$. If the spins are in equilibrium with lattice the distribution of the different spins over the magnetic levels is determined by the Boltzmann law

\begin{equation}\label{Boltzlaw}
    \frac{n_{-}}{n_{+}}=\exp\{\frac{\omega_0}{k_BT_L}\}
\end{equation}

where $T_L$ is a lattice temperature. Such an equilibrium can be violated, for example, after saturation by radio-frequency field. In this case one can describe spin system introducing spin temperature $T_s$   that is
distinguished from lattice temperature and is determined by
the expression

\begin{equation}\label{Boltzlaw1}
    \frac{n_{-}}{n_{+}}=\exp\{\frac{\omega_0}{k_BT_s}\}
\end{equation}

\begin{figure}[htb]	
\centering
\includegraphics[width=12cm]{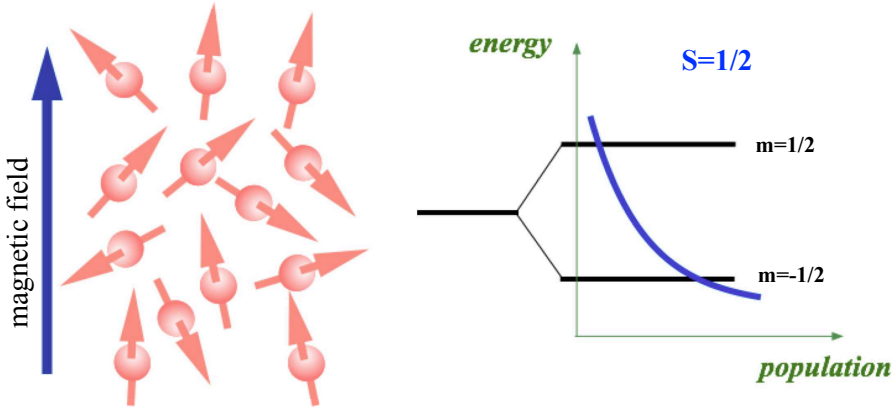}
\caption{Illustration of spin temperature concept for the case $S=1/2$. In the lower row the populations of level are shown for the case of positive, negative and infinite spin temperatures.}\label{spintemp}
\end{figure}

It should be noted that in the case of thermal equilibrium
with the lattice the levels of highest energy are always less populated than the lower ones. At the saturation it is possible to create situations in which the highest levels are more populated than the lower ones. If one assumes Equation \ref{Boltzlaw1} to be still valid this situation corresponds to negative spin temperatures. The spin system is not then of course in equilibrium with lattice because negative temperature is only defined for systems with an upper bound in the energy spectrum. The Figure \ref{spintemp} illustrates the concept of spin temperature for the case $S=1/2$ and positive, negative and infinite values of temperature.

In this part we shall discuss the thermodynamics of spin systems at low temperatures. We shall define "low temperature" as a temperature at which the energy of spin
in strong external magnetic field is of the same order as the average heat energy or exceeds it, i.e. the following condition is true:

\begin{equation}\label{lowtemp}
    \omega_0 \gtrsim k_BT_L
\end{equation}

In the case of localized electron spins in insulators this condition denotes a magnetic field of about 50 kG and a temperature about 1 K.
Concerning the energy of spin-spin interactions the assumption is made that it is small compared to the energy of spins in the external magnetic field. As we are not interested in the effects connected with phase transition into magnetically ordered state then we can assume that

\begin{equation}\label{lowtempSS}
    E_{ss} \ll k_BT_L
\end{equation}

where $E_{ss}$ is an average energy of spin-spin interactions. The temperatures at which an average heat energy of spins becomes more than the energy of a spin in an external field are called "high temperatures". The transition from the range of high temperatures into the range of low temperatures is accompanied first of all by the essential changes of the thermodynamic properties of spin system.

\subsection{High-temperature thermodynamics of spin system}
Let a consider the regular lattice of spins in an external magnetic field. The corresponding Hamiltonian of spin system has the following form

\begin{equation}\label{Hamspin}
   \mathcal{H}_s=\mathcal{H}_z+\mathcal{H}_{ss}
\end{equation}

where

\begin{equation}\label{HamspinZ}
   \mathcal{H}_z=\omega_0 \sum_{j}S_{j}^z
\end{equation}

is the Hamiltonian describing the Zeeman interaction of spins with the constant magnetic field directed along $z$-axis, $S_{j}^z$ is a longitudinal component of $j$-th spin. In Equation \ref{Hamspin}   $\mathcal{H}_{ss}$   describes the interactions of spins between themselves. For the sake of simplicity we will not take into account further interactions such as hyperfine interactions or interaction with crystal field. Temporarily we omit the interactions such as spin-phonon interactions and interactions with other external fields.
The energy levels of the Hamiltonian $\mathcal{H}_z$ are strongly degenerated because for any eigenvalue of the Hamiltonian one can find many combinations in which the eigenvalues of operators $S_{j}^z$ can be taken. This degeneracy is removed by the spin-spin interactions. Further we will suppose the external field to be large compared to the internal fields induced by spin-spin interactions. Therefore the Hamiltonian of spin-spin interactions $\mathcal{H}_{ss}$  can be considered as a perturbation. In the first order of the perturbation theory only those terms of $\mathcal{H}_{ss}$ which don't cause the change of any eigenvalue of the Hamiltonian $\mathcal{H}_{z}$ will give a contribution into the splitting of the energy level corresponding to this eigenvalue. So in the first order in the perturbation only part of $\mathcal{H}_{ss}$   that commutates with $\mathcal{H}_{z}$  will give a contribution in the broadening of the Zeeman levels. Usually this part is called the secular part and is presented as (4)

\begin{equation}\label{HamspinSS}
   \mathcal{H}_{ss}=\frac{1}{2} \sum_{i,j}(A_{ij}S_{i}^zS_{j}^z+B_{ij}S_{i}^{+}S_{j}^{-})
\end{equation}

where $S_{i}^{+}$ and $S_{i}^{-}$ are the transverse components of $j$-spin. Further we shall use the notation $\mathcal{H}_{ss}$ exactly for the
secular part of the Hamiltonian of spin-spin interactions. In this section we don't need the explicit form of spin-spin interaction constants $A_{ij}$  and $B_{ij}$  . Thus the energy spectrum of the Hamiltonian $\mathcal{H}_{s}$  represents quasi-continued equidistant bands of energy levels and these bands are separated in the energy by equal intervals $\omega_0$ .

The distribution of spins over these levels is the result of spin-spin interactions during some characteristic time $T_2$ that is of order of inverse width of the magnetic resonance line shape. As it follows from the experience usually in solids this time $T_2$ is significantly shorter than spin-lattice relaxation time $T_1$. The distribution of spins over the energy levels entirely determines the thermodynamic state of a spin system. It was appeared to be very convenient to introduce two spin temperatures for describing this distribution.

Formally the idea of two spin temperatures can appear from the following considerations. As long as spin system placed into strong external magnetic field has two first integrals - they are the Zeeman energy and the energy of the secular part of spin-spin interactions - then in order to describe a sufficiently slow evolution of any thermodynamic state it is natural to choose these first integrals as the thermodynamic coordinates. Such a choice becomes very attractive at high temperatures because the Zeeman energy and the energy of the secular part of spin-spin interaction are statistically independent in this case. The statistical independence denotes that one  can change a value of one energy without changing a value of the other, i.e. spin system is decoupled to two subsystems the energies of which are additive.

To explain the latter now we consider a quasi-equilibrium distribution, described by the following density matrix:

\begin{equation}\label{density}
\begin{array}{rcl}
  \rho_s=Q_{s}^{-1}\exp \{-\alpha \mathcal{H}_z-\beta \mathcal{H}_{ss}\}\\
  Q_{s}=Sp\exp \{-\alpha \mathcal{H}_z-\beta \mathcal{H}_{ss}\}
\end{array}
\end{equation}

Such a form of quasi-equilibrium distribution takes place due
to the fact of the availability of two invariants of motion.
In Equation \ref{density} parameters $\alpha$ and $\beta$ linked to the operators $\mathcal{H}_{z}$ and  $\mathcal{H}_{ss}$
are thermodynamically conjugative parameters for
the Zeeman energy and the energy of spin-spin interactions respectively. We can expand the exponent in Equation \ref{density} in powers of $\alpha \mathcal{H}_{z}$ and $\beta \mathcal{H}_{ss}$ and keep only the linear terms. As we shall see later such a linearization corresponds to the high temperature approximation. In the linear approximation in $\alpha \mathcal{H}_{z}$ and  $\beta \mathcal{H}_{ss}$,  the density matrix is reduced to

\begin{equation}\label{densitylin}
  \rho_s= \{1-\alpha \mathcal{H}_z-\beta \mathcal{H}_{ss}\}/Sp\hat{1}
 \end{equation}

Here $\hat{1}$ is the unit matrix. Therefore these expressions describe the expectation value of the Zeeman energy and the expectation value of the energy of spin-spin interactions:

\begin{equation}\label{expvalues}
\begin{array}{rcr}
  \langle \mathcal{H}_z \rangle = Sp\rho_s  \mathcal{H}_z \simeq -\alpha Sp \mathcal{H}_{z}^2/Sp\hat{1}\\
  \langle \mathcal{H}_{ss} \rangle = Sp\rho_s  \mathcal{H}_{ss} \simeq -\beta Sp \mathcal{H}_{ss}^2/Sp\hat{1}
\end{array}
\end{equation}

where we have taken into account the orthogonality of operators $\mathcal{H}_{z}$   and $\mathcal{H}_{ss}$   following from Equations \ref{HamspinZ} and \ref{HamspinSS}:

\begin{equation}\label{ortho}
  Sp \mathcal{H}_{z} \mathcal{H}_{ss}=0.
 \end{equation}

One can see from Equations \ref{expvalues} that in the linear approximation in
$\alpha \mathcal{H}_{z}$ and $\beta \mathcal{H}_{ss}$:

\begin{equation}\label{independ1}
  \frac{\partial \langle \mathcal{H}_{z} \rangle}{\partial \beta} =0.
 \end{equation}

On the other hand we have the explicit expression for this partial derivative :

\begin{equation}\label{independ2}
  \frac{\partial \langle \mathcal{H}_{z} \rangle}{\partial \beta} = \frac{\partial}{\partial \beta} \frac{Sp\exp\{-\alpha \mathcal{H}_{z} - \beta \mathcal{H}_{ss}\}\mathcal{H}_{z}}{Sp\exp\{-\alpha \mathcal{H}_{z} - \beta \mathcal{H}_{ss}\}}=-\langle \mathcal{H}_{z}  \mathcal{H}_{ss}\rangle+\langle \mathcal{H}_{z}\rangle \langle \mathcal{H}_{ss}\rangle
 \end{equation}

Comparing the right-hand sides of Equation \ref{independ1} and Eqquation \ref{independ2} we conclude that in the linear approximation in $\alpha \mathcal{H}_{z}$ and $\beta \mathcal{H}_{ss}$

\begin{equation}\label{independ3}
  \langle \mathcal{H}_{z}  \mathcal{H}_{ss}\rangle \simeq \langle \mathcal{H}_{z}\rangle \langle \mathcal{H}_{ss}\rangle
 \end{equation}

The Equation \ref{independ3} means the statistical independence of the first integrals. We remind the reader that two or more events are statistically independent if each individual event is not influenced by the occurrence of any other and Equation \ref{independ3} corresponds to the well-known rule of probabilities multiplying.

An entropy of the quasi-equilibrium distribution defined by
Equation \ref{densitylin} is equal in the linear approximation in $\alpha \mathcal{H}_{z}$ and $\beta \mathcal{H}_{ss}$ to

\begin{equation}\label{entropy}
  S=-k_BSp\rho_s \ln{\rho_s}\simeq-k_B\{-\ln{Sp\hat{1}}+\alpha^{2}Sp\mathcal{H}_{z}^{2}/Sp\hat{1}+\beta^2Sp\mathcal{H}_{ss}^2/Sp\hat{1}\}.
 \end{equation}

We see that the entropy is just a sum of contributions coming from
the Zeeman interaction and spin-spin interactions. Each of these contributions depends on $\alpha$ or $\beta$ only. The obvious identities follow from Equation \ref{entropy}:

\begin{equation}\label{temps}
\begin{array}{lcl}
  \alpha=\dfrac{1}{k_B}\dfrac{\partial S}{\partial \langle \mathcal{H}_{z}\rangle}=\dfrac{1}{k_BT_z}\\
  \beta=\dfrac{1}{k_B}\dfrac{\partial S}{\partial \langle \mathcal{H}_{ss}\rangle}=\dfrac{1}{k_BT_{ss}}
\end{array}
\end{equation}

where the parameters $T_z$ and $T_{ss}$ have the physical meaning of temperature of the Zeeman subsystem and temperature of the reservoir of spin-spin interactions respectively. The linear approximation in $\alpha \mathcal{H}_{z}$ and $\beta \mathcal{H}_{ss}$  corresponds to the situation when the energy of spin in external magnetic field and one in internal fields are small compared to the heat energy. Therefore such a linearization corresponds to the high temperature approximation. Further according to the tradition we shall call the parameters $\alpha$ and $\beta$ the temperatures of subsystems.

\begin{figure}[htb]	
\centering
\includegraphics[width=6cm]{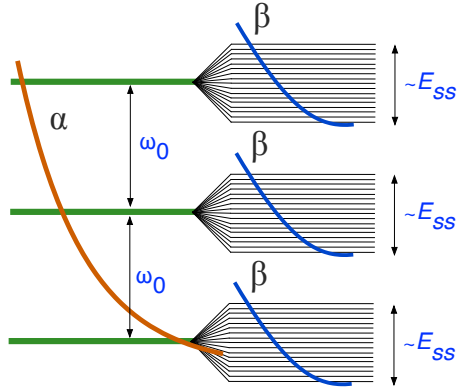}
\caption{The distribution of the spins over the energy levels of the spin system at high temperatures. It is possible to say about two spin temperatures - $\alpha$ and $\beta$ in high-temperature approximation}\label{2spin}
\end{figure}

In the thermal equilibrium with lattice the spin system is characterized by one single temperature (the temperatures  $\alpha$ and $\beta$ coincide). In the general case any thermodynamic state of spin system in high temperature approximation can be described by two temperatures: one for a Boltzmann distribution between the Zeeman levels, and a second one describing the ordering of the spins in the local fields induced by spin-spin interactions. A pictorial illustration is given in Figure \ref{2spin}.

So at high temperatures we have two statistically independent subsystems - the Zeeman subsystem and the reservoir of spin-spin interactions -  which are characterized by the
corresponding independent temperatures. It is the approximation in which the well-known Provotorov's theory was built \citep{prov62a,prov62b,prov63},. The developed two spin temperature formalism approach was appeared to be very fruitful to understand and to describe many problems of magnetic resonance, like saturation, spin-lattice relaxation etc. \citep{abragam94,goldman70,wolf79,atsarkin72}.

\subsection{Low-temperature thermodynamics of spin system}
At low temperatures the factorization condition ( see Equation \ref{independ3}) is violated because in the expanding of exponent one has to take high-order terms into account and the right-hand side of Equation \ref{independ1} will not be equal to zero. Consequently our subsystems became statistically dependent ( for more details the reader should refer to \citep{phil64,tayur89,tayur90} ). The advantage of the above-mentioned choice of thermodynamic coordinates is lost. Besides the Zeeman temperature doesn't have the physical meaning of temperature, and the temperature of the reservoir of spin-spin interactions is the real temperature of the whole system \citep{phil64}.

In order to be convinced of it we now note the following. As long as the Zeeman energy and the energy of spin-spin interactions are the first integrals then any linear combination of these energies with constant coefficients will be the first integral as well. In particular the total energy of spin system

\begin{equation}\label{energyS}
E_s=Sp\rho_s \mathcal{H}_{s}=Sp\rho_s \mathcal{H}_{z}+Sp\rho_s \mathcal{H}_{ss}=\langle \mathcal{H}_{zz}\rangle+\langle \mathcal{H}_{ss}\rangle
 \end{equation}

and the expectation value of $z$-component of the total spin

\begin{equation}\label{totalS}
\langle S\rangle=Sp\rho_s \sum_{j}S_{j}^z=\frac{1}{\omega_0}Sp\rho_s \mathcal{H}_{z}=\frac{1}{\omega_0}\langle \mathcal{H}_{z}\rangle
 \end{equation}

are the first integrals. If we choose these invariants as thermodynamic coordinates we should have an analogy of a thermodynamic system with fixed number of particles. The total number of particles and the $z$-component of total spin are the operators with discrete spectrum and these operators play a similar role. Therefore we can use a method analogous to that of the grand canonical ensemble and consider a Gibbs ensemble of systems characterized by the density matrix

\begin{equation}\label{densityG}
\begin{array}{rcl}
  \tilde{\rho}_s=\tilde{Q}_{s}^{-1}\exp \{-\beta \mathcal{H}_s+\nu S_z\}\\
  \tilde{Q}_{s}=Sp\exp \{-\beta \mathcal{H}_s+\nu S_z\},
\end{array}
\end{equation}

where the multiplier $\nu$ is fixed by the condition that $\langle S_z\rangle$ is given ( now $\langle \ldots \rangle$ denotes the averaging with the density matrix \ref{densityG} ).

The entropy of the distribution in Eq.[19] may be written as

\begin{equation}\label{entropyG}
  S=-k_BSp\tilde{\rho}_s \ln{\tilde{\rho}_s}=-k_B\{-\ln{\tilde{Q}}-\beta \langle \mathcal{H}_{s}\rangle+\nu \langle S_z\rangle \}.
 \end{equation}

Then taking into account the identity

\begin{equation}\label{identityG}
 \langle \mathcal{H}_{s}\rangle=\omega_0 \langle S_z\rangle+ \langle \mathcal{H}_{ss}\rangle,
 \end{equation}

which follows from the Equations \ref{Hamspin} and \ref{HamspinZ}  the entropy may be written as

\begin{equation}\label{entropyG1}
  S=k_B\beta  \langle \mathcal{H}_{ss}\rangle-k_B\langle S_z\rangle (\nu-\omega_0 \beta)+k_B \ln{\tilde{Q}}
 \end{equation}

One can use the definition of temperature for systems with
fixed number of particles and obtain from  Equations \ref{entropyG} and
\ref{entropyG1}

\begin{equation}\label{temperspin}
 \frac{1}{k_BT_s}=\left(\frac{\partial S}{\partial \langle \mathcal H_s\rangle}\right)_{\langle S^{z}\rangle}=\frac{S}{\partial \langle \mathcal H_{ss}\rangle}=\beta
 \end{equation}

\begin{figure}[htb]	
\centering
\includegraphics[width=6cm]{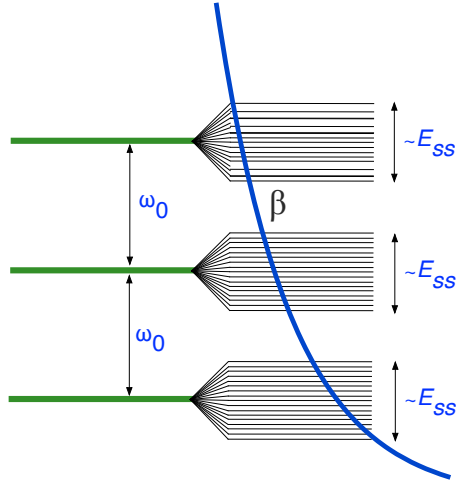}
\caption{The distribution of the spins over the energy levels of the spin system at low temperatures. Only the temperature of the reservoir of spin-spin interactions $\beta$ is the real temperature of spin system}\label{spintemplow}
\end{figure}

This means that the temperature of the reservoir of spin-spin interactions is the real temperature of spin system (see Figure \ref{spintemplow}). On the other hand one can write from Equation \ref{entropyG} the expression for the total energy

\begin{equation}\label{identityG1}
 \langle \mathcal{H}_{s}\rangle=\frac{1}{k_B\beta}S+\frac{\nu}{\beta}-\frac{1}{\beta}\ln{\tilde Q}
 \end{equation}

whence a simple relation between multiplier $\nu$ and the chemical potential follows

\begin{equation}\label{chemspin}
\mu=\left(\frac{\partial \langle \mathcal H_s\rangle}{\partial \langle S^{z}\rangle}\right)_{S}=\frac{\nu}{\beta}
 \end{equation}

(here we used the definition of the chemical potential for the grand canonical ensemble ). The state of the Zeeman subsystem is determined by the value of $z$-component of total spin or by the multiplier $\nu$ related to the chemical potential. Comparing the Equations \ref{density}, \ref{densityG} and \ref{chemspin} one gets a simple relation between the chemical potential of spin system and the parameters $\alpha$  and $\beta$:

\begin{equation}\label{chemspin1}
\mu=\omega_0\left(1-\frac{\alpha}{\beta}\right)
 \end{equation}

For further details concerning the using of chemical potential in spin thermodynamics the reader should refer to \citep{phil64}.
To calculate the expectation values of energies $\langle \mathcal H_z\rangle$ and $\langle \mathcal H_{ss}\rangle$ at low temperatures one has to take into account the higher-order terms in the expansion of the density matrix in Equation \ref{densitylin}. As a consequence the factorization condition \ref{ortho} is violated and the Zeeman subsystem and the reservoir of spin-spin interactions cannot be considered as independent. So the advantage of the above-mentioned choice of thermodynamic coordinates is lost. Besides at low temperatures the entropy written in terms $\alpha$  and $\beta$

\begin{equation}\label{entropyG2}
  S=\alpha  \langle \mathcal{H}_{z}\rangle+\beta  \langle \mathcal{H}_{ss}\rangle+k_B \ln{Q_S}
 \end{equation}

is not a sum of items each of which depends on $\alpha$ or $\beta$ only. Therefore at low temperatures the parameter a has no physical meaning of the inverse temperature of the Zeeman subsystem. In such a situation it is more plausible to use the expectation value of jz-component of total spin ( in other words the magnetization of spin system or the number of flipped spins) and the total energy as thermodynamic coordinates. Chemical potential and spin temperature are thermodynamically conjugated coordinates for them respectively. Such a choice of thermodynamic coordinates ( with taking into account the Equation \ref{chemspin1} ) has been made in \citep{haas80,haas81,tayur89,tayur90}. We note that one can also choose $\alpha$ or $\beta$ as the thermodynamic coordinates at low temperatures but it is necessary to take into account the correlation between them and the fact that parameter a has no physical meaning of the inverse temperature of the Zeeman subsystem.

So, by use very simple, model system - systems of interacting spins in an external magnetic field - we have demonstrated that the definition of temperature depends crucially on the assumptions made for this procedure. And one has to be very careful doing so.

\section{Conclusion}
The physical quantity ``temperature'' is a cornerstone of thermodynamics and statistical physics.  In the present paper the short introduction to the classical concept of temperature for macroscopic equilibrium systems was given. The concept of temperature was discussed regarding the nanoscale physics and non-extensive systems. It was shown forget that it is necessary to remember about the conditions to be satisfied in order to introduce ``temperature'' in macroscopic physics. The concept of ``spin temperature'' in condensed matter physics was reviewed and the advantages of thermodynamic approach to the problems of magnetism was illustrated. Partially, two temperatures spin thermodynamics was analyzed and the conditions when such approach is valid was studied.

%\bibliography{temp}

\end{document}